\documentclass[10pt,journal,compsoc]{IEEEtran}

\ifCLASSINFOpdf
\else
\fi

\usepackage{algorithmicx}
\usepackage{algpseudocode}
\usepackage{amsmath}
\usepackage{amssymb}
\usepackage{array}
\usepackage{booktabs}
\usepackage{caption}
\usepackage{cmap}
\usepackage{colortbl}
\usepackage{comment}
\usepackage{epstopdf}
\usepackage{etex}
\usepackage{fancyvrb}
\usepackage{fixltx2e}
\usepackage{float}
\usepackage{framed}
\usepackage{fp}
\usepackage{graphicx}
\usepackage{hhline}
\usepackage{hyperref}
\usepackage{lettrine}
\usepackage{lineno}
\usepackage{listings}
\usepackage{longtable}
\usepackage{makecell}
\usepackage{multirow}
\usepackage[numbers]{natbib}
\usepackage{pdfpages}
\usepackage{soul}
\usepackage{subcaption}
\usepackage{tabularx}
\usepackage[normalem]{ulem}
\usepackage{url} 
\usepackage{xcolor}
\usepackage{xpatch}

\definecolor{dkgreen}{rgb}{0,0.6,0}
\definecolor{gray}{rgb}{0.5,0.5,0.5}
\definecolor{mauve}{rgb}{0.58,0,0.82}
\definecolor{lgray}{rgb}{0.98,0.98,0.995}
\definecolor{darkslategray}{rgb}{0.18, 0.31, 0.31}
\definecolor{darksienna}{rgb}{0.24, 0.08, 0.08}
\definecolor{dimgray}{rgb}{0.41, 0.41, 0.41}
\definecolor{Gray}{gray}{0.9}

\begin{document}
\title{Mobile App Privacy in Software Engineering Research: A Systematic Mapping Study}

\author{Fahimeh~Ebrahimi, Miroslav~Tushev, and~Anas~Mahmoud
        }

\markboth{IEEE Transactions on Software Engineering,~Vol.~14, No.~8, October~2019}%
{Tushev \MakeLowercase{\textit{et al.}}: Modeling Linguistic Change in Open Source Software}

\IEEEtitleabstractindextext{
\begin{abstract}
Mobile applications (apps) have become deeply personal, constantly demanding access to privacy-sensitive information in exchange for more personalized user experiences. Such privacy-invading practices have generated major multidimensional and unconventional privacy concerns among app users. To address these concerns, the research on mobile app privacy has experienced rapid growth over the past decade. In general, this line of research is aimed at systematically exposing the privacy practices of apps and proposing solutions to protect the privacy of mobile app users. In this survey paper, we conduct a systematic mapping study of 54 Software Engineering (SE) primary studies on mobile app privacy. Our objectives are to \textbf{a)} explore trends in SE app privacy research, \textbf{b)} categorize existing evidence, and \textbf{c)} identify potential directions for future research. Our results show that existing literature can be divided into four main categories: privacy policy, requirements, user perspective, and leak detection. Furthermore, our survey reveals an imbalance between these categories---  majority of existing research focuses on proposing tools for detecting privacy leaks, with less studies targeting privacy requirements and policy and even less on user perspective. Finally, our survey exposes several gaps in existing research and suggests areas for improvement.    
\end{abstract}

\begin{IEEEkeywords}
Privacy, mobile application, systematic mapping study.
\end{IEEEkeywords}}

\IEEEpeerreviewmaketitle
\maketitle

\section{Introduction}
\lettrine{P}{rivacy} can be hard to define. The word itself is derived from the Latin \textit{Privatus}, which means \textit{``withdraw from public life''}. However, the majority of modern definitions can be traced back  Brandeis and Warren~\cite{Warren90} who defined privacy in 1890 as \textit{``the right to be let alone''}. This definition provided a basis for ensuring the legal protection of privacy as a fundamental human right. Since then, this definition has been revisited numerous times to count for the plethora of political, social, and technological advances in society. For instance, to meet the growing dimensions of the concept, Solove~\cite{Solove02} expanded the definition of privacy to include limited access to the self, secrecy, control of personal information, personhood, and intimacy. Another notable definition was introduced by Westin~\cite{Westin68}, who in 1968, described privacy as \textit{``the right to select what personal information about me is known to what people''} and later in 2003 as \textit{``the claim of an individual to determine what information about himself or herself should be known to others''}~\cite{Westin03}. Both definitions emphasized privacy as a control over personal information.  

The proliferation of mobile devices over the past decade along with their unique operational characteristics have imposed new challenges on end-users privacy. In general, mobile apps are designed with a set of goals in mind. A goal can be described as any desirable user objective that the system under consideration should achieve~\cite{Lamsweerde01}. However, driven by the fierce market competition, app developers often deviate from their original goals~\cite{Aydin17}. These deviations typically come in the form of extreme privacy-invading tactics, such as constant location-tracking~\cite{Li17}, unsolicited usage data collection~\cite{Aydin17,Wang18}, or any form of features that are engineered to lure users into sacrificing their privacy in exchange for more personalized services~\cite{Acquisti17,Papadopoulos17}. 

These intrusive, and sometimes borderline unethical, practices have led to the emergence of new and more significant privacy concerns among mobile app users. Such concerns often revolve around the types of information apps are asking for, who should or should not have access to this information, and how to prevent misuse of this access. In fact, these new challenges have prompted researchers to broaden existing definitions to include, in addition to personal information, device-specific information that can be used as identifiers, including installed apps, connected WIFI, operating system's build information, and carrier~\cite{Papadopoulos17}.  

In response to these challenges, the research on mobile app privacy has witnessed rapid growth over the past decade. In general, studies in this domain cover a broad range of topics, tackling privacy from a user, system, and even legal perspectives. To categorize, summarize, and synthesized this body of research, in this paper, we conduct a survey of existing Software Engineering (SE) literature related to mobile app privacy. Our survey takes the form of a Systematic Mapping Study (SMS). SMSs can be considered a form of Systematic Literature Reviews (SLRs), however, SMSs tend to be more qualitative in nature, mainly focused on the classification and thematic analysis of existing literature in a scientifically rigorous way. SLRs, on the other hand, are concerned with using statistical meta-analysis methods to synthesize the outcomes of empirical studies~\cite{Kitchenham09,Wohlin12}. SMSs have been long used in research as powerful tools for organizing and categorizing existing evidence and identifying areas for improvement~\cite{Kitchenham12}. A thorough discussion of the difference between SLRs and SMS is available in Kitchenham et al.~\cite{Kitchenham12} and Petersen et al.~\cite{Petersen08}. In general, our objectives in this survey paper include:  

\begin{itemize}
\item Providing descriptive statistics of existing primary studies on mobile app privacy published in SE venues between 2010 and 2018.
\item Systematically categorizing and sub-categorizing these primary studies based on the research problems they address. 
\item Summarizing existing evidence under each category and identifying research gaps and potential areas for improvement.  
\end{itemize}

The remainder of this paper is organized as follows. Section~\ref{method} describes our research method, including our research questions and primary study identification process. Section~\ref{summary} categorizes and summarizes existing primary studies and identifies directions of future work under each category. Section~\ref{discuss} discusses the main findings and limitations of our work. Finally, Section~\ref{conclude} concludes the paper.


\section{Method}
\label{method}
The survey presented in this paper takes the form of a Systematic Mapping Study. According to Kitchenham et al.~\cite{Kitchenham12}, a mapping study follows the same principled process as systematic literature reviews~\cite{Wohlin12}. This process consists of three main steps: planning, conducting, and reporting. Under the \textit{planning} phase, the need for the review is justified, the review protocol is established, and the research questions are defined. During the \textit{conducting} phase, the review protocol is put into action, including the identification of primary studies, categorizing, and synthesizing existing evidence. Finally, under the \textit{reporting} phase, the results are reported in a way that is tailored for the intended audience. In this section, we describe the main steps of our survey in detail.     

\subsection{Research questions}
It is essential to identify a set of research questions before taking on a review study. Research questions are necessary to set the \textit{scope} of the primary studies that should be considered in the search process and outline the objectives of the review. In this survey, our research questions are: 

\begin{itemize}
\item \textbf{RQ\textsubscript{1}}: \textbf{\textit{How much evidence is available in Software Engineering research about mobile app privacy?}}
Under this research question, we seek to determine the number of papers published on mobile app privacy in Software Engineering research. Our objective is to systematically assess the community's effort on this problem. Research venues tackling mobile privacy from other perspectives such Machine Learning or Computer Human Interaction, are excluded.    

\item \textbf{RQ\textsubscript{1}}: \textbf{\textit{What are the main categories of existing research?}}
Under this question, we seek to build a classification scheme of existing primary studies~\cite{Petersen08}. Our objective is to narrow down the subjects of mobile app privacy research to few well-defined areas. Such categorization can provide researchers and practitioners with a much needed reference structure, or visual map, that can be easily navigated to get a quick overview of the frequency of publications and their main trends over the past decade. 

\item \textbf{RQ\textsubscript{3}}: \textbf{\textit{What are the limitations of the current research?}}
Under this question, we seek to identify areas for improvement in existing SE mobile app privacy research. Specifically, through our categorization, we seek to identify areas where the research has not progressed in comparison to other areas. Furthermore, we analyze the future work sections of exiting papers to outline the authors' visions of future work directions in their areas of research. Our objective is to help junior researchers, or researchers breaking into the field, to identify the areas which are lacking evidence, or the views of experts on research directions that need more attention or worth pursuing.  
\end{itemize}

\subsection{Search Process: Identifying primary studies}
Our scope includes Software Engineering primary studies (papers that are not surveys, or secondary studies) tackling privacy of mobile apps, published in the period from 2010 to 2018. To identify these papers, we followed a three-step process. At the first step, we searched four scientific databases: Google Scholar, ACM Digital Library, IEEE Xplore, and arXiv. Our search query can be described as follows: 

\begin{framed}
\noindent \textit{(Mobile \scalebox{0.88}{\textbf{AND}} (app \scalebox{0.88}{\textbf{OR}} apps) \scalebox{0.88}{\textbf{AND}} privacy) \scalebox{0.88}{\textbf{OR}} (mobile \scalebox{0.88}{\textbf{AND}} application \scalebox{0.88}{\textbf{AND}} privacy) \scalebox{0.88}{\textbf{OR}} (mobile \scalebox{0.88}{\textbf{AND}} app \scalebox{0.88}{\textbf{AND}} privacy \scalebox{0.88}{\textbf{AND}} permission) \scalebox{0.88}{\textbf{OR}} (mobile \scalebox{0.88}{\textbf{AND}} app \scalebox{0.88}{\textbf{AND}} privacy \scalebox{0.88}{\textbf{AND}} (policy \scalebox{0.88}{\textbf{OR}} policies)) \scalebox{0.88}{\textbf{OR}} (privacy \scalebox{0.88}{\textbf{AND}} app store)}
\end{framed}  

%

The number of papers identified at the first step was 524. However, since we are only considering software engineering research, we only considered papers published in SE venues. Limiting the search to SE venues was necessary for two main reasons. First, privacy is a very broad concept, as of June 2019, a trivial search for \textit{mobile privacy} on Google Scholar returns 3,810,000 hits. Therefore, it is important to narrow down the scope of search in order to get a reasonable number of primary studies. Second, our target audience is the SE community, or researchers and practitioners who typically publish in SE venues. To identify our venues, we relied on previous SMS and SLR studies in the literature~\cite{Kitchenham09SLR}. In general, any venue with \textit{``Software Engineering''} in its name, or any venue (workshop) that was held with a major SE venue were included in our analysis. Furthermore, we enforced the time scope of 2010 - 2018 on selected papers. 2010 was the year where initial work on mobile app privacy started emerging as a new line of research in SE, following the lunch of the first app store in 2008. A search for primary studies before 2010 did not return any hits. In summary, 85 papers were found in SE venues between 2010 and 2018. These venues, along with the number of papers found in each, are shown in Table.~\ref{Tab:Venues}, the number of papers identified based on each of our sub-queries are shown in Table. 2. 

\begin{table*}
\scriptsize
\centering
\renewcommand{\arraystretch}{1.2}
\caption{Selected journals and conference proceedings.}
\label{Tab:Venues}
\begin{tabular*}{59em}{l|c|c|c}
\bottomrule
\Xhline{1.5\arrayrulewidth}
{\textbf{Source}} & \textbf{Acronym}    & \textbf{Count}   &  \textbf{Included}    \\ \hline
ACM Transactions on Software Engineering and Methodology							&	TOSEM	&	1	&	1	\\ 
Communications of the ACM												&	CACM	&	1	&	0	\\ 
Empirical Software Engineering												&	EMSE	&	3	&	2	\\ 
IEEE Software							      					        		&	IEEE SW	&	1	&	1	\\ 
IEEE Transactions on Software Engineering			      							&      TSE		&	5	&	5	\\ 
Information and Software Technology				     							&      IST		&	2	&	1	\\ 
International Conference Fundamental Approaches to Software Engineering				&	FASE		&	1	&	1	\\
Inter. Conference on Automated Software Engineering								&	ASE		&	5	&	6	\\ 
Inter. Conference on Mobile Software Engineering and Systems						&    MOBILESoft &	15	&	6	\\
Inter. Conference on Software Engineering										&	ICSE		&	14	&	11	\\ 
Inter. Conference on Software Engineering Companion								&    ICSE Comp. &	2	&	0	\\ 
Inter. Conference Software Analysis, Evolution, and Reengineering						&	SANER	&	4	&	2	\\
Inter. Journal of Secure Software Engineering									&	IJSSE	&	1	&	0	\\
Inter. Requirements Engineering Conference										&	RE 		&	6	&	5	\\ 
Inter. Symposium on Dependable Software Engineering: Theories, Tools, and Applications	&	SETTA	&	1	&	1	\\
Inter. Symposium on Software Reliability Engineering								&	ISSRE	&	5	&	3	\\
Inter. Symposium on Software Testing and Analysis								&	ISSTA	&	5	&	4	\\ 
Inter. Symposium on Foundations of Software Engineering							&	FSE		&	4	&	1	\\ 
Inter. Working Conference on Requirements Engineering: Foundation for Software Quality    	&	REFSQ	&	1	&	0	\\
Inter. Workshop on Automation of Software Test									&	AST		&	1	&	1	\\
Inter. Workshop on Requirements Engineering and Law								&	RELAW	&	1	&	1	\\
Journal of Systems and Software												&	JSS		&	3	&	0	\\ 
Software: Practice and Experience											&	SPE		&	3	&	2	\\ 
Workshop on Requirements Engineering										&	WRE		&	0	&	0	\\
\hline
\textbf{Sum} 															&			& \textbf{85}&\textbf{54}\\
 \hline
\end{tabular*}
\end{table*}

\begin{figure}\centering
\includegraphics[trim={1.5cm 6.5cm 10cm 16.5cm},clip, width=9.6cm]{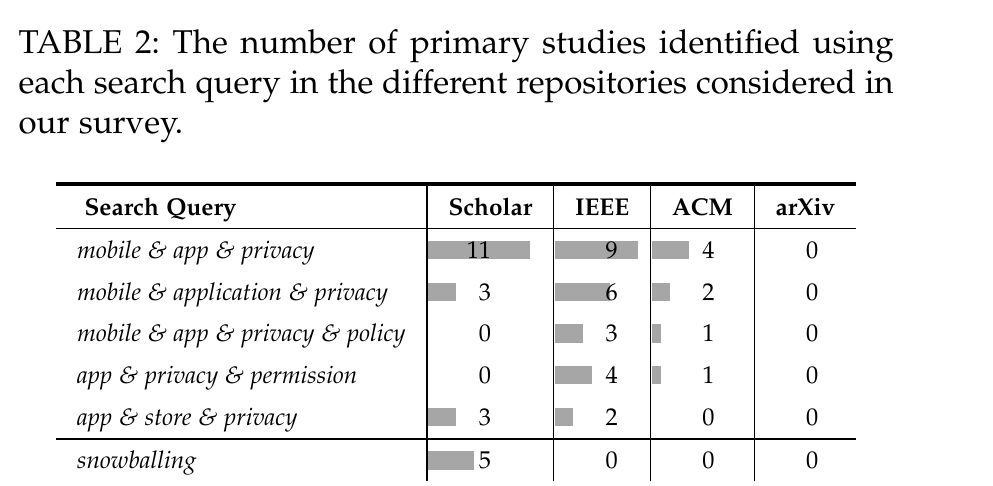}
\label{partial}
\end{figure}

\subsection{Inclusion and exclusion criteria}
In SMSs and SRLs, the inclusion and exclusion criteria are used as a basis for selecting primary studies. Such criteria should be determined beforehand during the planning phase. Our inclusion criteria in this paper are:

\begin{itemize}
\item Books, papers, technical reports, and grey literature.  
\item The study explicitly addresses privacy aspects of mobile apps. 
\item The study is published in English.
\end{itemize}

We used the following \textit{exclusion criteria} to exclude any studies that were irrelevant to our survey goals:

\begin{itemize}
\item Short papers (less than 4 pages), editorials, and summaries of keynote, or tutorial papers.
\item Secondary studies (i.e., existing SLRs or SMSs).
\item Duplicate reports of the same study. In case of duplication, the most recent version is selected.  
\end{itemize}

To include and exclude papers, each paper was examined by each of the three authors individually. Specifically, each author read the title, abstract, and body of each of our 85 papers to determine its relevance to our survey. Each judge flagged each paper as \texttt{Include}, \texttt{Neutral}, and \texttt{Exclude}. The paper was then included or excluded based on the protocol shown in Table~\ref{Tab:RelevanceDecisionMaking}. In total, 36 papers out of our 85 papers were excluded from the analysis. 

To reduce the risk of omitting relevant studies, we also performed a lightweight backward-snowballing on the included papers~\cite{Wohlin14}. We basically inspected the studies cited by each of our included primary studies and the publications that subsequently cited the study, using Google Scholar. In total, 6 more papers were identified, raising the number of primary studies to be included in our survey to 54 papers. The histogram in Fig. 1 shows the number of identified studies per year from 2010 to 2018. In this histogram, studies found using snowballing were counted as Google Scholar papers. A summary of our search process is shown in the schematic diagram in Fig.~\ref{UDDD} and the distribution of final set of included papers over venue type is shown in Fig. 3. A detailed description of our search process can be found in (\url{http://seel.cse.lsu.edu/data/TSE2019.xlsx})

\begin{figure}\centering
\includegraphics[trim={1.5cm 12cm 10cm 10.6cm},clip, width=9cm]{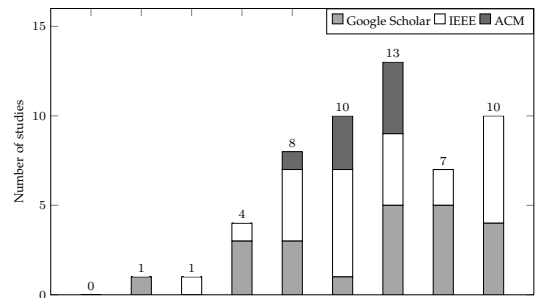}
\label{fig:StudyPerYear}
\caption{The number of published papers per year from 2010 to 2018.}
\end{figure}

\begin{figure}\centering
\includegraphics[scale=1]{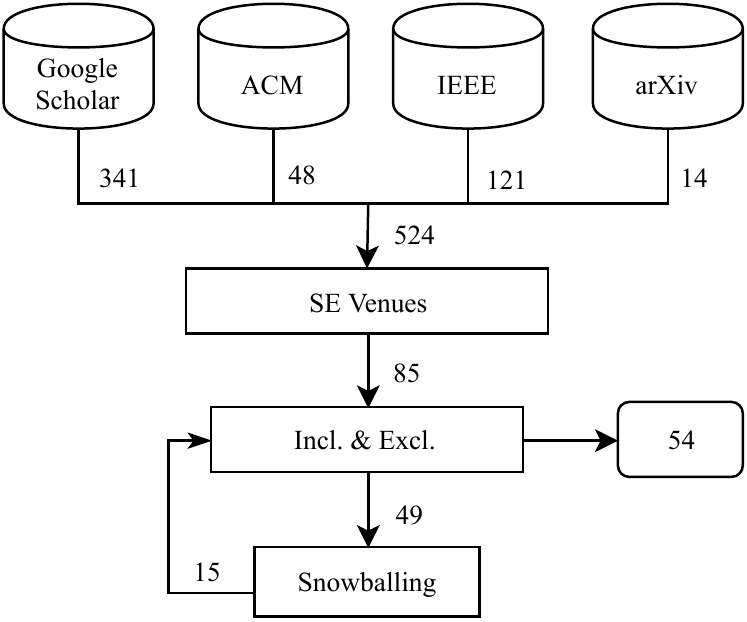}
\caption{A schematic diagram of our proposed research plan.}
\label{UDDD}
\end{figure}

\begin{figure}\centering
\includegraphics[trim={1.5cm 6.6cm 10cm 17.1cm},clip, width=9.6cm]{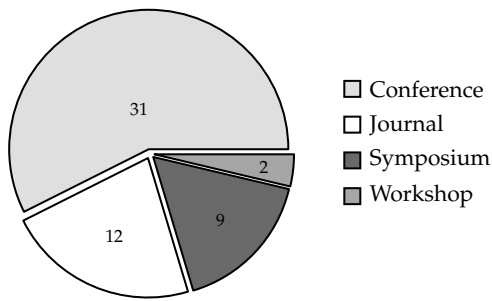}
\label{fig:type}
\caption{The distribution of primary study over venue type.}
\end{figure} 

\section{Primary Study Categorization}
\label{summary}
To develop our classification scheme (\textit{RQ\textsubscript{2}}), we follow the protocol proposed by Petersen et al.~\cite{Petersen08}. In general, this protocol can be described as follows: 
\begin{itemize}
\item The reviewers read the abstracts and look for keywords and concepts that reflect the contribution of the paper. If the abstract is poorly structured, the body of the paper is examined.  
\item A set of keywords representative of the context, or contribution of the paper, are extracted.   
\item The paper is assigned to an existing category based on its keywords. If the category does not exist, a new category is created.   
\end{itemize}

In general, we identified four different categories: privacy policy, privacy requirements, mobile app users' perspective of privacy, and privacy leak analysis. The last category was further classified into two main categories of static and mixed static-dynamic methods. In what follows, we summarize the papers under each category and we highlight any areas for improvement.


\subsection{Privacy Policy}
Mobile apps' privacy policies act as contracts between app users and providers. In general, policies inform users about how their personal information is collected, used, shared, and protected by the app~\cite{Wang18,Bhatia16}. Problems in this domain arise from inconsistencies between privacy claims in the policy and the actual behavior of the app. These inconsistencies (also known as violations) often take the form of omission errors, where the policy fails to notify the user about a specific information-privacy practice~\cite{Wang18,Young11}. In what follows, we summarize the papers classified under this category. 

\begin{table}
\scriptsize
\centering
\renewcommand{\arraystretch}{1.5}
\caption{The decision making process. IN: Include, EX: Exclude, and NU: Neutral.}
\label{Tab:RelevanceDecisionMaking}
\begin{tabular}{l|l|l|l}
\bottomrule
\Xhline{1.5\arrayrulewidth}
\textbf{Researcher\textsubscript{1}} & \textbf{Researcher\textsubscript{2}}  & \textbf{Researcher\textsubscript{3}}   & \textbf{Final Decision}    \\ \hline

IN  &      IN  &   {IN/EX/NU }  &  Include \\ 
EX  &      EX  &   {IN/EX/NU }  &  Exclude \\ 
NU  &      NU  &   {IN/EX/NU }  & Consensus Meeting \\ 
IN  &      EX  &   NU  &   Consensus Meeting \\ 

\bottomrule
\end{tabular}
\end{table}

\subsubsection{Existing Work}
Yu et al.~\cite{Yu18,Yu16} introduced \texttt{TAPVerifier}, an automated approach for establishing semantic correlations between apps' privacy policies and their bytecode. The authors' goal was to identify apps' description-to-behavior fidelity, or whether the app behaves as advertised. \texttt{TAPVerifier} relied on text analysis to automatically establish and match the semantic meaning of the app's privacy policy, its bytecode, app store description, and permissions. Evaluating \texttt{TAPVerifier} on a dataset of 1,200 apps revealed that privacy policies were more likely to describe privacy-related behaviors than descriptions. The results also showed that the proposed approach was able to remove up to 59.4\% of false alerts generated by tools such as \texttt{Whyper} and \texttt{AutoCog}.    

Slavin et al.~\cite{Slavin16} proposed a semi-automated framework to help app developers detect inconsistencies between their privacy policies and code. The proposed framework linked policy phrases to Application Program Interface (API) methods that produced sensitive information. Information flow analysis was then used to detect misalignments by identifying methods that sent data to third-party remote servers. The proposed framework was empirically evaluated on 477 Android apps and discovered~341 potential privacy policy violations.

Aydin et al.~\cite{Aydin17} introduced \texttt{VisiDroid}, a visual configuration interface for end users to understand and configure the way their private information was utilized. The objective was to tailor privacy policies according to the preferences and sensitivities of the user. The authors targeted privacy threats emerging from contextual advertising. Specifically, through a special GUI, end users were able to configure constraints on the context. These configurations were then enforced by the app by anonymizing private fields before being sent to remote servers. The usability of \texttt{VisiDroid} was evaluated using a user study of 20 participants. All participants were able to finish all their privacy configuration tasks successfully.  

Wang et al.~\cite{Wang18} proposed an approach to automatically detect app privacy policy violations based on user input. The proposed approach mapped each Graphical User Interface (GUI) input element of the app to ontology concepts, which in turn were matched with text from the privacy policy. Static information flow analysis was then used to detect inconsistencies between the data collection behavior of the app and the statements of data-collection in the policy (i.e. \textit{privacy leaks}). Evaluating the proposed approach over 120 popular apps, sampled from the finance, health, and dating domains, that tit was able to detect 39 strong and weak violations in the studied apps.

Yu et al.~\cite{Yu18-2} presented \texttt{PPCkecker}, a tool for assessing the trustworthiness of Android apps' privacy policies. \texttt{PPChecker} employed natural-language processing (NLP) techniques, such as syntactic analysis, to analyze privacy policies, and adopted static program analysis approaches to inspect if the app's code collected, retained, or disclosed personal information. Based on this analysis, the authors identified five kinds of problems in privacy policies: incompleteness, incorrectness, impreciseness, inconsistency, and not being user-friendly. Applying \texttt{PPCkecker} to 2,500 popular apps revealed that 1,850 apps (i.e., 74.0\%) had at least one kind of the identified privacy problems.

\subsubsection{Summary and Future Work}
Our survey revealed that privacy policy research revolves around detecting policy violations by mapping privacy claims in the app's privacy policy to information flow in the app's code. Static code analysis tools such as FlowDroid~\cite{Arzt14} are typically used to track information flow in bytecode. Furthermore, majority of the papers under this category propose some sort of a working prototype to implement their findings (e.g., \texttt{PPCkecker}~\cite{Yu18-2} and \texttt{TAPVerifier}~\cite{Yu18} and \texttt{VisiDroid}~\cite{Aydin17}) and evaluation is mainly carried out on Andriod apps as it is possible to get their APK files. Finally, our survey revealed that a large percentage of apps were either non-compliant with their stated privacy claims, did not provide a policy, or provided a policy that was ambiguous or incomprehensible to average users~\cite{Zimmeck16}. 

In general, the line of research under this category is still at its preliminary stages~\cite{Slavin16}. Specific directions of future work include combining more dynamic and static analysis methods with policy analysis methods~\cite{Yu18-2}, developing techniques to further trace the flow of user information to the respective organizations or servers~\cite{Slavin16}, and expanding the scope of privacy threats to areas other than contextual adverting~\cite{Aydin17,Wang18}. In terms of tool support, future policy analysis tools should be engineered to meet the demands of their specific userbase (e.g., developers, users, regulators). Such tools should help app developers and policy authors to stay aware of any violations in their apps, enable end-users to determine the trustworthiness of the app~\cite{Yu18-2}, and help regulators, such as the U.S. Federal Trade Commission (FTC), or even app stores, to identify questionable or malicious apps~\cite{Slavin16}. 

\subsection{Privacy in Mobile App User Feedback}
App stores provide a mechanism for app users to express their opinions about apps in the form of textual reviews and meta-data (e.g., star ratings). Recent research revealed that around 40\% of app store reviews on popular app stores contain actionable maintenance requests~\cite{Iacob13,Jha19}. Addressing these requests was found to have a significant positive impact on the rating of the app~\cite{Noei19,Palomba15}. Under this category, we summarize papers that identified privacy as a major concern in user feedback.  

\subsubsection{Existing Work}
Khalid et al.~\cite{Khalid15} conducted an analytical study of user reviews to help developers better anticipate and prioritize possible user complaints. The authors manually examined and classified thousands of one and two star app reviews, sampled from 20 iOS apps. The analysis uncovered 12 types of common users complaints. Analyzing the impact of each compliant type on the apps' ratings revealed that reviews complaining about privacy invading business practices were often associated with the most negative impact on the app as reflected in the ratings.

Mcllroy et al.~\cite{McIlroy16} conducted a qualitative analysis of close to 7,000 user reviews sampled from Google Play and the Apple App Store. The results showed that a substantial amount (30\%) of reviews raised more than one technical issue. In terms of privacy, the authors found that 17\% of reviews raised some sort of privacy concerns also raised other types of functional or technical concerns. The concentration of these concerns seemed to vary among app categories, where the Shopping category had the highest percentage of privacy complaints. The authors also reported that privacy issues were expressed using more varied language than other technical issues. 

Ciurumelea et al.~\cite{Ciurumelea17} proposed a taxonomy to analyze reviews and codes of mobile apps for a better release planning. The authors defined specific categories of user reviews that were highly relevant for developers during software maintenance, including compatibility, usage, resources, pricing, and privacy. A machine learning based prototype was then introduced to classify reviews and recommend source code files that were likely to be modified to handle issues raised in the reviews. Evaluating the proposed approach over 39 Android apps showed that it was able to detect privacy and protection issues with up to 83\% precision and 96\% recall. 

Scoccia et al.~\cite{Scoccia18} conducted a large-scale empirical study to investigate end users perception of the new Android's run-time permission system. The authors collected over 4.3 million user reviews from 5,572 Google Play apps. A sample of 3,574 permission-related reviews were analyzed using machine learning (Naive Bayes and Support Vector Machines) and NLP techniques. The most common permission-related complaints were then classified into a taxonomy of issues. The results showed that 8\% of collected apps had reviews with negative comments about permissions. These permission-related reviews occurred in almost all app categories.

\subsubsection{Summary and Future Work}
Our survey revealed that end-user privacy concerns, while not as common as other functional issues (reliability and usability), are widespread among almost all app categories and can have a significant negative impact on the ratings of apps~\cite{McIlroy16,Khalid15}. However, our survey exposed a gap in existing research related to mining end-user privacy concerns. Specifically, the majority of existing work proposes generic approaches for mining user feedback for generic maintenance requests (bug reports and feature requests)~\cite{Martin17}. These techniques often struggle when it comes to detecting specific user issues~\cite{Jha19}. This emphasizes the need for data mining techniques that are specifically tailored for detecting privacy concerns in user feedback, for instance, building taxonomies of user-defined privacy concerns across the different categories of the app store. Extracted information should be then used to facilitate a privacy-aware app design and development process.   

\subsection{Mobile App Privacy Requirements}
Under this category, we summarize SE primary studies related to mining, specifying, and enforcing privacy requirements of mobile apps.     

\subsubsection{Existing Work}
Young~\cite{Young11} proposed an approach to obtain software requirements from privacy policies. The author used a grounded theory approach to systematically analyze policies and extract commitments, privileges, and rights conveyed within the policies. Extracted information was classified into 12 different procedural (the privacy practice) and legal (the law enforces the practice) categories and then operationalized into software requirements using predefined requirements templates. The analysis was conducted using a case study on 17 health-care privacy policies. The results showed that the proposed approach achieved better coverage in comparison to legal-based approaches since polices often describe procedural practices rather than legal practices. 

Tun et al.~\cite{Tun12} introduced privacy arguments as a means of analyzing privacy requirements in general and selective disclosure requirements in particular. These arguments enabled users to express their personal privacy preferences through an extended argumentation language. User preferences were then used to represent highly dynamic selective disclosure requirements, and to relate them to the software architecture in order to enable system adaptation to runtime privacy requirements. The practical feasibility of the proposed approach was demonstrated over a mobile app that was developed by the authors.   

Omoronyia et al.~\cite{Omoronyia13} proposed an adaptive privacy framework to support the selective disclosure of personal information in mobile apps. The framework exploited privacy awareness requirements (PAR) to identify the runtime privacy properties that should be satisfy in order to manage the changing user privacy concerns. The framework was evaluated over multiple cases where the failures and satisfactions of PAR were realized. The results showed that applications that failed to satisfy PAR were unable to regulate information flow based on the utility of disclosure, while applications that satisfied PAR were able regulate the disclosure of information with changing context. 

Thomas et al.~\cite{Thomas14} proposed a problem analysis framework to extract and refine privacy requirements for mobile apps from data gathered via empirical end-user studies. The framework provided analytical tools such as thematic coding, heuristics, facet questions, and extraction rules to enable a structured analysis of the various privacy problems experienced by users. Furthermore, the framework provided a privacy language to reason about and model privacy requirements. The operation of the framework was demonstrated using data collected through a case study targeting users of the Facebook app. The results showed that the approach was able to provide systematic aid to software engineers in deriving privacy requirements that addressed end-users' privacy concerns and needs.

Mai et al.~\cite{Mai18} proposed a use-case driven method to support the specification of security and privacy requirements of multi-device software ecosystems, including mobile and wearable device applications. Specifically, the authors integrated an existing approach of modeling security threats, their mitigation, and their relations to use cases in a \textit{misuse} case diagram. The authors further introduced templates for specifying mitigation schemes and misuse case specifications. A regular expression engine was then used to automatically report inconsistencies among artifacts and between the templates and specifications. The proposed approach was applied over industrial multi-device healthcare project and was further evaluated using structured interviews with four software engineers. The proposed approach was able to precisely specify and analyze security threats along with threat scenarios and their mitigations.

Breaux et al.~\cite{Breaux14,Breaux13} introduced \texttt{Eddy}, a methodology to map a well-defined subset of privacy requirements from natural language text to a formal language in description logic. \texttt{Eddy} was intended to help developers detect conflicting privacy requirements within a policy and enable the tracing of data flows within these policies. A computer simulation was used to demonstrate the scalability of the proposed language toolset to specifications of reasonable size. In a more recent work, Breaux et al.~\cite{Breaux15} targeted the privacy risks which arise from the increased information sharing across services (APIs). Specifically, the authors presented a new \texttt{Eddy} extension to formally model multi-party data flows requirements. These models were aligned with three critical privacy principles: purpose specification, collection limitation, and use limitation. The proposed approach was evaluated through a case study on the mobile app Waze and three of its service providers: Facebook, Amazon Web Services, and Flurry.com. The results showed that the proposed extensions were able to detect conflicts between Waze and Flurry privacy polices as well as violations of the principle. In addition, the analysis uncovered two privacy specification design patterns whereby specification authors can bypass the limitation principle. 

Van Der Sype and Maalej~\cite{Van14} proposed a set of concrete requirements for app developers to protect users' privacy, focusing on the disclosure of users' personal data to third parties. These requirements were derived from the EU Data Protection Directive (Directive 95/46/EC) principles for fair and lawful processing. These principles were categorized into four groups: purpose limitation, data minimization, data security, and transparency. The authors further discussed the tension between law and technology, providing concrete strategies for developers to ease this tension.  

Liu and Simpson~\cite{Liu16} introduced \texttt{PPTMA} (Privacy-Preserving Targeted Mobile Advertising), a solution to achieve a balance between user privacy and utility in the context of mobile ads. The authors established the functional and non-functional requirements, design guidelines, and architecture of \texttt{PPTMA} based on the unique privacy challenges of mobile advertising. \texttt{PPTMA} was implemented as a background service which can be dynamically configured to enable users to adjust their privacy settings to control ads' access to their private information. The prototype was evaluated on 200 Android apps. The results showed that \texttt{PTTMA} was able to operate with low memory and power consumption, providing effective means for ad-scanning. 

\subsubsection{Summary and Future Work}
Our survey revealed that specifying privacy requirements for mobile apps can be a very challenging process. These requirements are highly dynamic and often context specific, depending on many factors such as time, location, information content, domain, and user preferences~\cite{Thomas14,Tun12,Omoronyia13}. This problem becomes even more challenging in multi-device ecosystems where the app has to communicate information with other devices over the network~\cite{Mai18}.    

In terms of future work, our survey exposed several potential areas of improvement in mobile privacy requirements research. These areas are mainly focused on devising practical techniques for eliciting, modeling, and managing mobile app privacy requirements. Elicitation techniques should take into account the constantly changing nature of privacy requirements under the different usage scenarios of the apps. This emphasizes the need for new ethnographic analysis techniques, such as interaction analyses, to obtain tacit knowledge of users' behavior in different contexts~\cite{Thomas14}. These techniques should also consider factors such as existing privacy regulations and privacy policies when generating the privacy requirements of the app~\cite{Van14,Young11}. 

Once privacy requirements are specified, modeling techniques should be proposed to represent the relations (synergies and trade-offs) of these requirements to other entities in their specific domain, including user goals and functional features. Existing functional and non-functional requirements modeling techniques, such as Feature Oriented Domain Analysis (FODA)~\cite{kang90} and Softgoal Interdependency Graphs (SIGs)~\cite{Bruno04,Eric09} can be used to show the mandatory, alternative, and optional features of the domain along with their commonalities and variabilities~\cite{Pohl05}. Once these models are generated, they can act as blueprints of privacy-aware requirements models that app developers in different application domains can use as references for their design. Furthermore, such models should be adaptive in order to reflect the constantly changing privacy concerns and the newly emerging threats~\cite{Omoronyia13}. Finally, the authors of primary studies emphasize the need for effective automated tool support and extrinsic evaluation strategies to empirically assess the performance of the proposed techniques in realistic settings (e.g., validation through experience)~\cite{Thomas14,Tun12,Liu16,Breaux15}.  

\subsection{Privacy Leak Analysis}
Under this category, we summarize papers that target privacy leaks and malicious behavior in mobile apps. In total, 33 primary studies were classified under this category. To better understand the summaries of these studies, we provide the following definitions:   

\begin{itemize}
\item \textbf{Static Analysis:} static analysis methods examine source code or its binary representations without executing the code. Such methods often employ control flow graph analysis~\cite{Li17}. FlowDroid~\cite{Arzt14} and IccTA~\cite{Li15Iccta} are examples of static analysis tools that are commonly used in privacy leak analysis.  

\item \textbf{Dynamic Analysis:} refers to monitoring the code as it executes, thus enabling a precise security analysis based on the run-time behavior of the software. In general, dynamic methods use data follow methods to track data from untrusted sources as the program executes to detect malicious behaviors~\cite{Newsome05}. 

\item \textbf{Taint Analysis:} is a static/dynamic code analysis method that is commonly used for privacy leak detection. The goal of the process is to observe and track tainted/untainted information flow between sources and sinks. In the context of mobile apps, a tainted source is any system call that access private data and a sink is all the possible ways that make data leave the system (e.g., leak)~\cite{Demissie16}. \texttt{TaintDroid}~\cite{Enck14} is a commonly used tool for taint analysis in Android apps.  

\item \textbf{Privacy Leak:} a privacy leak could be defined as any path that sends user's sensitive information to the outside world (e.g., a third party server) without the user's permission~\cite{Zhang17}. 

\item \textbf{Permission:} In the context of mobile apps, a permission governs the type of user information the app is allowed to access. This could range from the data stored in the mobile device's memory to any peripherals of the device like the camera or the motion sensor. Users must grant permissions to the app before it can access these resources.

\end{itemize}

Given the large number of papers classified under this category, we combine related papers summaries into two subcategories of purely static, and mixed static/dynamic method analysis. 

\subsubsection{Static Analysis Methods}
Feng et al.~\cite{Feng14} presented \texttt{Apposcopy}, a semantic-based method to detect apps that leak private user information. \texttt{Apposcopy} used a high-level language to describe the semantic signatures of malware families and static analysis to determine whether an app matches a malware signature. Signature matching was enabled through static taint analysis and a novel code representation called Inter-Component Call Graph. These graphs tracked information flow using pointer analysis. \texttt{Apposcopy} was evaluated on 1027 malicious and 11,215 benign Android apps. The results showed that it detected malware with high accuracy (90\%) and that its signatures were resilient to various program obfuscations.

Huang et al.~\cite{Huang14} introduced \texttt{AsDroid}, a technique for detecting stealthy malicious behaviors in Android apps. The technique consisted of two parts: a static analysis module for extracting API invocations, and a UI analysis module for extracting the displayed in the interface. Any mismatch between the two components' outputs was considered as a potential stealthy behavior. A dataset of 182 potentially problematic Android apps was used to evaluate \texttt{AsDroid}. \texttt{AsDroid} was able to report stealthy behaviors in 113 apps, with 28 false positives and 11 false negatives.

Bartel et al.~\cite{Bartel14} demonstrated that off-the-shelf static analysis was insufficient for extracting mappings between API methods and permissions in Android apps. To overcome these limitations, the authors proposed a set of static analysis techniques. Three of these techniques were based on class hierarchy, including string analysis, service identity inversion, and entry point. Two other techniques: service redirection and service initialization, leveraged a field-sensitive module, called \texttt{Spark}. Evaluation of the proposed methods over 1,421 Android apps revealed that 9\% of the apps were accessed more permissions than necessary. 

Shen et al.~\cite{Shen14} and Holavanalli et al.\cite{Holavanalli13} proposed Flow Permissions, a technique to enrich the Android permission mechanism with flow analysis. Flow Permissions was implemented through \texttt{BlueSeal}, a tool that employed static analysis to extract information flow between permission domains within an app or between apps via IPC (Inter Process Communication) mechanisms. The cross-app permission analysis compared information flows within new apps to those of already-installed apps to derive new Flow Permissions. The effectiveness and utility of Flow Permissions was evaluated using a dataset of 2,992 popular Android apps and 1,047 malicious Android apps and a survey of 540 participants. The results indicated that Flow Permissions was practical to deploy, was able to significantly impact user's decisions to install an app and provided visibility into the holistic behavior of mobile apps.

Gorla~\cite{Gorla14} proposed \texttt{CHABADA}, a technique for identifying inconsistencies between advertised behavior of Android apps and their implemented behavior. \texttt{CHABADA} performed topic modeling on app descriptions and then clustered apps using k-means. Within each cluster, sensitive APIs which were governed by a user permission were identified. Outliers with respect to API usage were detected using One-Class Support Vectors Machine (SVM). These outliers were considered potentially malicious activity. \texttt{CHABADA} was applied on a set of 22,500+ Android apps. The prototype was able to detect several anomalies and flag 56\% of novel malware. 

Xiao et al.~\cite{Xiao15} proposed a user-aware privacy control mechanism to track how sensitive data was used inside mobile apps. The authors employed static analysis to extract information flows inside apps. The computed information flows were classified into safe and unsafe flows based on a tamper analysis that tracks
whether private data is obscured before escaping through output channels. To evaluate the effectiveness of the proposed method, 50 users opinions were collected, out of which ~90\% considered the method to be effective to protect their privacy. The results also showed that the proposed approach could be used to automatically provide default privacy settings for apps. 

Huang et al.~\cite{Huang15} introduced \texttt{Dflow}, a context-sensitive information flow type system, and \texttt{DroidInfer}, the corresponding type inference analysis for detecting privacy leaks in Android apps. \texttt{DroidInfer} improved the scalability of \texttt{FlowDroid} by avoiding points-to-based static analysis. The proposed approach was a type-based taint analysis technique which explained information flows using context-free language. Evaluating \texttt{DroidInfer} on 144 free Android apps showed that it was able to detect privacy leaks in Android apps with a false-positive rate of 15.7\%.

Yang et al.~\cite{Yang15Appcontext} introduced \texttt{AppContext}, a mechanism that statically analyzed sensitive behaviors of Android apps. \texttt{AppContext} extracted the contexts (events and conditions) that triggered security-sensitive behaviors. The authors applied SVM to classify an app as malware or benign based on the extracted contexts of the app. \texttt{AppContext} was evaluated on a dataset of 202 malicious and 633 benign Android apps. The results showed that \texttt{AppContext} was able to correctly identify malicious apps with 87.7\% precision and 95\% recall.

Li et al.~\cite{Li15Iccta} presented \texttt{IccTA}, a static taint analysis tool for detecting inter-component privacy leaks in Android apps (paths from sources to sinks). \texttt{IccTA} relied on propagating context information among components. Specifically, \texttt{IccTA} built a control-flow graph of the whole Android application. This allowed propagating the context between Android components and yielding a highly precise data-flow analysis. \texttt{IccTA} adequately modeled the lifecycle and callback methods to detect ICC (Inter-component communication) based privacy leaks. Evaluating \texttt{IccTA} revealed that it outperformed existing tools (e.g., FlowDroid) on two popular benchmark datasets, DroidBench and ICC-Bench. Furthermore, it was able to detect 534 ICC  leaks in 108 apps from the \texttt{MalGenome} dataset of malware and 2,395 ICC leaks in 337 apps in a set of 15,000 Google Play apps.

Bagheri et al.~\cite{Bagheri15} presented \texttt{COVERT}, a tool for compositional analysis of Android inter-app vulnerabilities. The authors utilized static code analysis to automatically model apps while they were installed, removed, or updated. A formal analyzer was used to match the extracted model with a set of vulnerability patterns to detect leakages occurred due to the interaction of apps comprising a system. \texttt{COVERT} was examined against privilege escalation, one of the most prominent inter-app vulnerabilities, using a dataset of 500 Android apps. The results showed that it was able to detect inter-app vulnerabilities in several bundles of popular apps.

Avdiienko et al.~\cite{Avdiienko15} introduced \texttt{MUDFLOW}, a tool for detecting malicious apps based on their treatment of sensitive data. \texttt{MUDFLOW} was trained on the data flow of benign apps (the most popular 2,866 Google Play apps). For every sensitive data source, the sensitive APIs to which this data flows was determined. This information was then used to automatically flag apps with suspicious features. Specifically, an app's information flow was extracted using FlowDroid~\cite{Arzt14}. \texttt{MUDFLOW} tagged an app as malicious if its data flow from sensitive sources was different from benign apps data flow. Applying \texttt{MUDFLOW} on a dataset of 25,577 malicious apps and 2,950 benign apps showed that it was able to correctly identify 86.4\% of all malware, and 90.1\% of malware leaking sensitive data.

Lee et al.~\cite{Lee16} presented \texttt{HybriDroid}, a static analysis framework for Android hybrid apps, or apps supporting multiple mobile platforms. The proposed framework analyzed inter-communications between Android Java and JavaScript. The authors relied on extensive testing of source code of apps to understand the semantics of interoperation mechanisms of Java and JavaScript and built call graphs for both Java and JavaScript via an on-the-fly pointer analysis. To demonstrate the types of analysis supported by the framework, the authors implemented a bug detector that identified programmer errors due to hybrid semantics, and a taint analyzer that found information leaks across language boundaries. These tools were empirically evaluated using 88 apps. The results provided an evidence on the practicality of the tools and their ability to detect previously uncovered bugs and privacy leaks in Android hybrid apps. 

Li et al.~\cite{Li16, Li16Droidra} proposed \texttt{DroidRA}, an instrumentation-based approach to tackle reflection in Android apps by propagating string constants. The authors applied code instrumentation to replace reflection calls with standard Java calls. \texttt{DroidRA} resolved string values in a context- and flow-sensitive manner. \texttt{DroidRA} was evaluated on apps from the \texttt{Droidbench} benchmark and a dataset of real-world apps. The results showed  that the approach was useful in uncovering dangerous code (e.g., sensitive API calls, sensitive data leaks), outperforming several state-of-the-art analyzers.   

Meng et al.~\cite{Meng16} proposed a semantic model for Android malware detection and classification. The model represented elements of malicious behaviors in malware variants as Deterministic Symbolic Automata with system APIs as transitions. Based on this model, the authors implemented an automatic malware analysis framework. The proposed framework was evaluated on a malware benchmark and a dataset of 223,170 real-world Android apps. The results showed that the framework outperformed several state-of-the-art malware detection approaches and anti-virus tools.

Rahman et al.~\cite{Rahman17} used 21 static code metrics, commonly used for defect prediction (e.g., bad coding practice and duplication), to assess the security and privacy risks associated with Android apps. The authors applied four statistical learners to predict the level of privacy risk: Radial-Basis SVM (r-SVM), CART (Classification and Regression Trees), k-Nearest Neighbor (KNN), and Random Forests (RF). Evaluation was carried out over 4,416 Android apps. The results showed that r-SVM was able to predict privacy risks level (e.g., high, medium, and low) with 83\% precision. 

Narayanan et al.~\cite{Narayanan18} introduced \texttt{MKLDroid}, a static analysis framework for malware detection and malicious code localization in Android apps. The framework systematically integrated multiple views (malware detection approaches such as sensitive APIs, system calls, and control-flow) of apps using a graph kernel to capture structural and contextual information from apps' dependency graphs. Multiple Kernel Learning (MKL) was then used to find the weighted combination of views which achieved the best detection accuracy. \texttt{MKLDroid} was evaluated through large-scale experiments on multiple datasets. The results showed that it outperformed multiple baselines in terms of accuracy while maintaining comparable efficiency. Furthermore, \texttt{MKLDroid} was able to identify malice classes with 94\% average recall.

Zhang et al.~\cite{Zhang18} developed \texttt{Ripple}, a static reflection analysis that tackled incomplete information environments of Android apps. Undetermined intents, behavior-unknown libraries, unmodeled services, and unresolved built-in containers were considered as incomplete information environments. \texttt{Ripple} applied type inference to resolve reflective calls, even if some data flow were missing. \texttt{Ripple} was validated over 17 Android apps with incomplete data-flows. The results showed that it was able to discover reflective targets with a low false positive rate of 21.9\%.

Shan et al.~\cite{Shan18} proposed an approach to support the characterization and detection of self-hiding behaviors in Android apps. The authors grouped these behaviors into three main categories: hiding an app's presence, blocking or removing traces of remote communications, and subverting the device's reminders. A suit of static analysis methods were employed to detect the defined self-hiding behaviors. Based on the detected behaviors, apps were labeled as benign or malicious. The proposed approach was evaluated on a data set of 9,452 Android apps. The results showed that each malicious app employed 1.5 self-hiding behaviors, on average. Furthermore, the proposed technique attained 87.19\% F-measure in detecting malicious apps.

Dilhara et al.~\cite{Dilhara18} presented \texttt{ARPDroid}, a technique to repair incompatible use of permissions in Android apps. \texttt{ARPDroid} performed static control flow analysis to extract program points where permission usages must be checked. Based on the results of the analysis, \texttt{ARPDroid} detected incompatible permission uses. All incompatibilities were then resolved using bytecode transformation. Evaluation of \texttt{ARPDroid} over 23 Android apps showed that it achieved 100\% detection accuracy and 90\% repair success rate.

Garcia et al.~\cite{Garcia18} introduced \texttt{RevealDroid}, a lightweight machine-learning-based approach for detecting malicious Android apps with privacy leaks and identifying their general families. \texttt{RevealDroid} leveraged API-based features (e.g., API method invocations), reflective invocations, and the internal behaviors of apps' native binaries (e.g., external calls made by native binaries). The authors used a linear SVM for malware detection and a CART classifier for family identification. \texttt{RevealDroid} was evaluated on a dataset of around 54,000 malicious and benign Android apps. The results showed 98\% accuracy in detecting malware and 95\% in determination of their families.

Canfora et al.~\cite{Canfora18} proposed \texttt{LEILA}, a tool for Android malware families detection and localizing malicious classes into the infected apps. \texttt{LEILA} applied model checking, using Milner's Calculus of Communicating Systems, to analyze and verify the Java Bytecode produced when the app's source code was compiled. This gave the tool an advantage in case the code was not available due to obfuscation. \texttt{LEILA} was evaluated on a dataset of malware families released from 2011 to 2016. The results showed accuracy levels between 97\% and 100\%. The authors also devised a set of rules for specifying malicious behaviors of kinds of malware families.

\subsubsection{Mixed Analysis Methods}
Xu et al.~\cite{Xu13} presented \texttt{Premlyzer}, a hybrid framework for automatically analyzing the uses of requested permissions in Android apps. The framework used call stack based analysis to identify where and how a permission is used in the app by analyzing the API calls that triggered the permission check. \texttt{Premlyzer} was evaluated using 51 malware/spyware families and over 110,000 Android apps. The results showed that \texttt{Permlyzer} was able to provide detailed permission use analysis and discover the characteristics of the permission uses in benign and malicious apps. 

Hay et al.~\cite{Hay15} proposed \texttt{IntentDroid}, a comprehensive testing algorithm for Android Inter-Application Communication (IAC) vulnerabilities. \texttt{IAC} enabled the reuse of functionality across apps via message passing, thus it could be used by malicious apps to infect benign apps. The authors started by describing 8 concrete vulnerability types, along with attack scenarios, that could result from the unsafe handling of \texttt{IAC} messages. To test an app, \texttt{IntentDroid} monitored any APIs responsible for security-relevant functionality and access to \texttt{IAC} data in order to recover \texttt{IAC}-relevant app-level behaviors. All the possible attack scenarios were then implemented to detect malicious activities. \texttt{IntentDroid} was evaluated over a dataset of 80 Android apps. The result revealed 150 IAC vulnerabilities with a recall rate of 92\%.  

Keng~\cite{Keng16} proposed a hybrid static/dynamic analysis to detect user-triggered causes and paths of privacy leaks in Android apps. The proposed approach used static analysis to capture activity transitional paths to identify data leakage paths. Technically, the author utilized callback control flow analysis to generate an activity transition graph. This allowed automated testing to directly traverse apps towards privacy related activities for verification. The user-triggered causes of privacy leaks were then displayed to users in the form of messages. The proposed approach was evaluated through a 2-week long field user-study with 47 Android users. The results showed that the generated privacy messages were effective for users in managing their privacy. 

In a more recent work, Keng et al.~\cite{Keng16Graph} introduced \texttt{MAMBA}, a new automated app testing technique that was used to extract test cases from control-flow graphs to simulate privacy-sensitive API calls. \texttt{MAMBA} initially searched for user events in control-flow graphs of callbacks generated from static analysis of app bytecode. Based on the extracted paths, the tool built test cases using user events that may trigger access to privacy-sensitive data in the apps. \texttt{MAMBA} then simulated the execution of the app using the generated test cases to detect any runtime leaks. Evaluation of \texttt{MAMBA} on 24 apps showed that \texttt{MAMBA}'s graph-aided directed testing was faster than an exhaustive testing approach.    

Lee et al.~\cite{Lee17} proposed \texttt{SEALANT}, a tool for analyzing Android inter-app vulnerabilities. \texttt{SEALANT} applied static code analysis to analyze architectural information of a given set of apps from their APKs.  The tool then used data-flow analysis and ICC pattern matching to detect vulnerable ICC where inter-app attacks could be launched. Extracted information was then visualized in integrated and compositional representations. Evaluating \texttt{SEALANT} over multiple datasets of Android apps showed that it was able to accurately identify vulnerable ICC paths and successfully aid users' understanding of inter-app vulnerabilities.

Zhang and Feng~\cite{Zhang17} proposed \texttt{AndroidLeaker}, a hybrid analysis tool for detecting privacy leaks in Android apps based on taint analysis. \texttt{AndroidLeaker} used static analysis (data flow analysis on call graphs) to detect privacy leaks in individual apps and dynamic analysis to prevent any leaks caused by cooperation of multiple apps. The goal was to reduce runtime overhead false alarms often associated with static and dynamic methods. \texttt{AndroidLeaker} was tested on 82 examples from \texttt{DroidBench}, a test suite released with \texttt{FlowDroid}. The results showed lower precision (82\%) but higher recall (69\%) than \texttt{FlowDroid}.

Xu et al.~\cite{Xu15} introduced \texttt{SpyAware}, a privacy leakage detection framework which examined the correlation between the data flow of privacy leakage and an app's execution path. \texttt{SpyAware} used a profiler module to profile app executions in binder calls and system calls. A feature extractor module was implemented to extract feature vectors of suspicious profile. The authors utilized SVM and Naive Bayes (NB) to train and predict \texttt{Spyware} executions based on the feature vectors. The proposed framework was evaluated over 100 popular Android apps. Experimental results showed 67.7\% accuracy in device ID leakage detection and 78.4\% accuracy in location-leakage detection. 

Demissie et al.~\cite{Demissie16} introduced Android Wicked Delegation (AWiDe), a dangerous case of delegation in which benign apps unintentionally exposed their access to sensitive resources to attackers (i.e., permissions are re-delegated to a malicious app). The authors formulated this problem as a problem of inadequate input validation of inter-app communication in Android. The authors compared static and dynamic approaches to automatically detect inadequate message validation. The first approach used static taint analysis to detect if the values used in privileged actions depended on potentially malicious data. The second approach used dynamic analysis with automatically generated input values to detect when data during execution was used in privileged actions by malicious apps. The proposed approach was validated on a dataset of 329 Android apps. The results showed that around 15 popular apps were affected by AWiDe vulnerabilities.

Moussa et al.~\cite{Moussa17} proposed \texttt{ACCUSE}, an approach to help users to assess the risk of installing an Android app. \texttt{ACCUSE} assigned an app a risk factor based on the Android classification of permissions (normal, dangerous, and system). The authors rescaled the risk factors with the app's rating and the number of downloads. \texttt{ACCUSE} visualized the total risk of installing an app before installation. Evaluation of \texttt{ACCUSE} on 12,576 Android apps suggested that it outperformed several existing malware detection tools.

Zhao et al.~\cite{Zhao18} proposed a system to assess whether an app was vulnerable to location inference attacks. Specifically, the authors employed a series of automatic testing mechanisms including UI match (dynamic testing) and API analysis (static analysis) to extract the location information an app provided. These apps were then classified into two categories based on the type of attacks: with and without distance limitations. Evaluating the proposed system on a dataset of 800 android apps revealed that 34.7\% of the passing apps are vulnerable location-leak attacks.





\subsubsection{Summary and Future Work}
Our survey revealed that majority of work under this category of primary studies is focused on introducing new and more effective methods for privacy leak detection. These methods often take the form of a working tool or a prototype that employs static, dynamic, or mixed models of code analysis to detect components of the apps (e.g., code, API calls, GUI components) that might lead to information leakage. 

Our survey also revealed that static code analysis methods are the most common, with fewer number of papers applying dynamic and mixed dynamic-static methods. In fact, even methods applying dynamic analysis often apply some sort of static analysis to complement the approach. This can be attributed to the fact that static methods are typically cheaper to execute and can generate more in depth information about data flow within the app's code. However, there seems to be a consensus that dynamic methods that track the execution traces of apps would give a far more detailed view of apps' behavior~\cite{Gorla14}. Finally, our survey of papers under this category revealed three main directions of future work. These directions can be identified as follows:

\begin{itemize}
\item \textbf{Improving accuracy:} the majority of papers suggest improving the proposed methods' accuracy as a main direction of future work. Accuracy, in general, is related to the ability of the proposed method to capture privacy leakage, or identify malware apps, or malicious behaviors with decent levels of precision and recall~\cite{Feng14,Yang15Appcontext,Lee16,Xu15}. Suggestions for improving accuracy involve incorporate dynamic analysis ~\cite{Xiao15,Avdiienko15,Bagheri15,Narayanan18,Zhang18,Dilhara18}, testing the proposed approach on larger datasets of apps or different benchmarks~\cite{Yang15Appcontext}, or validating the effectiveness of the results using human studies~\cite{Demissie16}.  

\item \textbf{Coverage:} another common line of future work aims to broaden the scope of research to investigate other types of attack models~\cite{Zhao18,Hay15}, including inter-app attacks~\cite{Lee17}, malware families~\cite{Canfora18}, malicious behaviors~\cite{Narayanan18}, and anomalies within specific application domains~\cite{Avdiienko15}.   

\item \textbf{Supporting other platforms:} evaluation in the majority of existing papers is carried over datasets of Android apps. Therefore, supporting other operating systems, mainly iOS, is identified in several studies as a main direction of future work~\cite{Hay15,Zhao18,Xiao15,Bartel14}. 
\end{itemize}

\section{Discussion}\label{discuss}
In this section, we provide more quantitative analysis results, describe our study limitations, and describe existing related work. 

\subsection{Quantitative summary}
\begin{figure}\centering
\includegraphics[trim={3.2cm 18.8cm 1cm 2.5cm},clip, width=9.5cm]{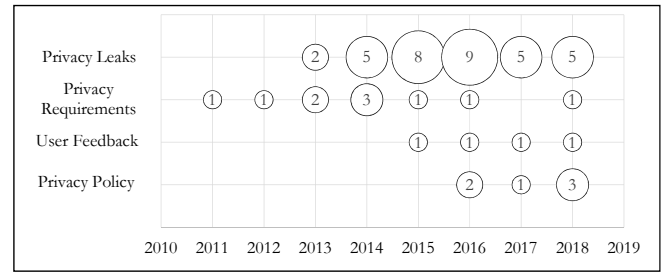}
\caption{A Bubble chart of the growth of the different categories over time.}
\label{BubblePlot}
\end{figure}

Our survey revealed that privacy leak analysis, using static and dynamic code analysis methods, is the largest category of primary studies. The categories of privacy requirements and policy are relatively comparable in size. However, they have significantly less primary studies than the privacy leak category. The category of user feedback has the smallest number of primary studies. Table~\ref{Tab:CategoriesTable} summarizes our categories and lists the primary studies classified under each category, and the Bubble chart in Fig~\ref{BubblePlot} shows the growth of these different categories (number of papers) over time. 

A common theme that our survey revealed was that the majority of primary studies proposed some sort of a tool or a working prototype to implement their approach or evaluate their findings. We list these tools along with their descriptions in Table~\ref{Tab:ToolsTable}. Another observation is that, except for few studies on requirements and policy~\cite{McIlroy16,Khalid15}, evaluation in majority of the primary studies was carried out over datasets of Android apps. This emphasizes the need for a community-wide effort to prepare benchmark datasets that support iOS apps.


\begin{table*}
\footnotesize 
\centering
\renewcommand{\arraystretch}{1.2}
\caption{Categories and subcategories of primary studies as identified in our survey.}
\label{Tab:CategoriesTable}
\smallskip 
\begin{tabular*}{\textwidth}{l|l|m{27em}|m{17em}}
\bottomrule
\Xhline{1.5\arrayrulewidth}

 \multicolumn{2}{l|}{\textbf{Category}}    & \textbf{ Description} & \textbf{Studies}  \\
  \hline

  \multicolumn{2}{l|}{Privacy Policy}   & Studies under this category target inconsistencies between apps claimed privacy practices, listed in the privacy policy, and their actual behavior. &   
\cite{Aydin17}, \cite{Wang18}, \cite{Yu18}, \cite{Yu16}, \cite{Slavin16}, \cite{Yu18-2}   \\
  \hline
  
 \multicolumn{2}{l|}{User Feedback}   &  Studies under this category are aimed at mining user privacy concerns from app store reviews. & \cite{Khalid15}, \cite{McIlroy16}, \cite{Ciurumelea17}, \cite{Scoccia18} \\
  \hline
  
  \multicolumn{2}{l|}{Privacy Requirements}      & Studies under this category are focused on mining, specifying, and enforcing privacy requirements of mobile apps. &  
 \cite{Young11}, \cite{Mai18}, \cite{Tun12}, \cite{Omoronyia13}, \cite{Thomas14}, \cite{Breaux14}, \cite{Breaux13}, \cite{Breaux15}, \cite{Liu16}, \cite{Van14}\\
  \hline

  \multirow{4}*{Privacy Leaks}  & Static Analysis & Studies in this category use static program analysis to detect privacy leakage in mobile apps. & 
\cite{Li15Iccta}, \cite{Feng14}, \cite{Huang14}, \cite{Bartel14}, \cite{Shen14}, \cite{Holavanalli13}, \cite{Gorla14}, \cite{Xiao15}, \cite{Huang15}, \cite{Yang15Appcontext}, \cite{Bagheri15}, \cite{Avdiienko15}, \cite{Lee16}, \cite{Li16}, \cite{Li16Droidra}, \cite{Meng16}, \cite{Rahman17}, \cite{Narayanan18}, \cite{Zhang18}, \cite{Shan18}, \cite{Dilhara18}, \cite{Garcia18}, \cite{Canfora18} \\ 
\cline{2-4}

 & Mixed Methods& These studies  combine static and dynamic methods for detecting privacy leaks. & 
\cite{Demissie16}, \cite{Zhang17},  \cite{Xu13}, \cite{Hay15}, \cite{Keng16}, \cite{Keng16Graph}, \cite{Lee17}, \cite{Xu15},  \cite{Moussa17}, \cite{Zhao18}\\
\bottomrule
\end{tabular*}
\end{table*}

\begin{table*}
\footnotesize 
\centering
\renewcommand{\arraystretch}{1.2}
\caption{The list of the tools and working prototype proposed in the primary studies summarized in our survey.}
\label{Tab:ToolsTable}
\smallskip 
\begin{tabular*}{\textwidth}{@{\extracolsep{\fill}}lll}
\bottomrule
\Xhline{1.5\arrayrulewidth}

{\textbf{Tool}} & \textbf{Category}  & \textbf{Brief Description}      \\ \hline

\texttt{TAPVerifier}~\cite{Yu18} &  Privacy Policy  & An application verification tool that utilizes privacy policies to detect privacy violations.\\ 
\texttt{VisiDroid}~\cite{Aydin17} & Privacy Policy  & A visual configuration interface that allows users to configure their privacy preferences.\\
\texttt{PPCkecker}~\cite{Yu18-2} & Privacy Policy  &  A tool for assessing the trustworthiness of the Android apps' privacy policies.\\ 

\texttt{PPTMA}~\cite{Liu16} & Privacy Requirements & A service to achieve a balance between user privacy and utility.\\

\texttt{Apposcopy}~\cite{Feng14} &  Static Analysis  &  A semantic-based static taint analysis method to detect malicious apps. \\
\texttt{AsDroid}~\cite{Huang14} & Static Analysis   &  A privacy leak detector that detects inconsistencies between API invocations and UI.\\
\texttt{BlueSeal}~\cite{Shen14, Holavanalli13} &  Static Analysis  & An automatic system to generate Flow Permissions in Android apps.\\
\texttt{CHABADA}~\cite{Gorla14} &  Static Analysis  & A technique to identify inconsistencies between apps' descriptions and their behavior.\\ 
\texttt{Dflow} + \texttt{DroidInfer}~\cite{Huang15} &  Static Analysis  & A context-sensitive information flow type system for detecting privacy leaks in Android apps. \\
\texttt{AppContext}~\cite{Yang15Appcontext} &  Static Analysis  & A privacy violation detection technique based on the context information. \\
\texttt{IccTA}~\cite{Li15Iccta} &  Static Analysis  & A static taint analyzer for detecting inter-component privacy leaks. \\
\texttt{COVERT}~\cite{Bagheri15} & Static Analysis  & A compositional analysis tool to detect Android inter-app vulnerabilities. \\ 
\texttt{MUDFLOW}~\cite{Avdiienko15} &  Static Analysis  & A static taint analysis tool for distinguishing malicious apps based on their data flow. \\
\texttt{HybriDroid}~\cite{Lee16}&  Static Analysis   &  A static analysis framework for hybrid apps.  \\
\texttt{DroidRA}~\cite{Li16, Li16Droidra} &  Static Analysis  & A string inference analysis for resolving reflection. \\
\texttt{MKLDroid}~\cite{Narayanan18} &  Static Analysis &  A context-aware, multi-view malware detection tool using graph kernels. \\ 
\texttt{Ripple}~\cite{Zhang18} &  Static Analysis  &  a reflection analysis that tackles incomplete information environments of Android apps. \\ 
\texttt{ARPDroid}~\cite{Dilhara18} & Static Analysis & A static control flow technique for detecting incompatible permission uses.\\
\texttt{RevealDroid}~\cite{Garcia18} &   Static Analysis  &  A lightweight machine-learning-based approach for detecting malicious apps.\\ 
\texttt{LEILA}~\cite{Canfora18} &  Static Analysis  & A Java bytecode analyzer exploiting model checking to detect malicious activities.\\ 

\texttt{Premlyzer}~\cite{Xu13} &  Mixed Methods   & A hybrid permission analysis tool for Android apps. \\
\texttt{IntentDroid}~\cite{Hay15} &  Mixed Methods   & A testing approach to detect Android inter-application communication vulnerabilities. \\
\texttt{MAMBA}~\cite{Keng16Graph} & Mixed Methods  &  A graph-aided directed testing system for the automated checking of privacy behaviors.\\
\texttt{SEALANT}~\cite{Lee17} &  Mixed Methods   & A tool for automated analysis and visualization of Android inter-app vulnerabilities. \\
\texttt{AndroidLeaker}~\cite{Zhang17} &  Mixed Methods   & A hybrid analysis tool for detecting privacy leaks in Android apps. \\
\texttt{SpyAware}~\cite{Xu15} & Mixed Methods & A privacy leakage detection framework based on data flow and execution path analysis.\\
\texttt{ACCUSE}~\cite{Moussa17} & Mixed Methods  & An approach to assess apps' installation risks.\\

\bottomrule
\end{tabular*}
\end{table*}

\subsection{Limitations}
In terms of limitations, we notice that some of our identified categories were more well-defined than others. For example, several of the primary studies under the user feedback category were not primarily intended to detect privacy concerns, rather they considered privacy as a user concern that could be extracted from user feedback. Furthermore, there were some studies that could be classified under multiple categories. For example, the primary study in~\cite{Young11} described techniques for extracting privacy requirements from policy. Therefore, it could be classified under both categories. However, since the main focus of the study is deriving privacy requirements, it was summarized under the requirements category. 

Other limitations might stem from our inclusion/exclusion protocol. Specifically, we relied on our own assessment of the literature in order to include or exclude papers. To mitigate this threat, we applied a systematic coding protocol, which included individual coding of primary studies and majority voting. We further acknowledge the fact that the quality of summaries might vary across the different categories and sometimes within the same category. We tried to mitigate this threat by imposing a standard structure on our summaries. Finally, our suggested directions of future work, while mainly relied on analyzing the future work sections of our primary studies,  were also based on our own assessment of the literature. This might raise some subjectivity concerns as experts in these specific fields might have different views of which specific problems should be perused, or which problems have more value to the research community. 

\subsection{Related surveys}
Since our study takes the form of a systematic survey, other related surveys, or secondary studies, were excluded from our survey. For instance, we identified two mobile app privacy surveys that appeared in SE venues, Sadeghi et al.~\cite{Sadeghi17} proposed a taxonomy and qualitative comparison of program analysis techniques for security assessment of Android apps, and Li et al.~\cite{Li17} conducted a systematic literature review of static analysis methods of Android apps. 

In general, our mapping study can be distinguished from these existing surveys along two main dimensions. First, we considered the problem from a SE perspective by only considering venues that are commonly known to be SE focused venues, and second, we extended our survey beyond code analysis to include a broader range of SE topics, such as policy, requirements, and user feedback. To that extent, our mapping study bridges an important gap in existing secondary literature on mobile apps privacy. 



\section{Conclusions}\label{conclude}
We conducted a systematic mapping study of Software Engineering literature on mobile apps privacy. Our objectives were to categorize and summarize the state-of-the-art and enumerate the challenges to be addressed by the Software Engineering research community. Our survey included 54 primary studies published in software engineering venues in the period between 2010 and 2018. These studies were classified, following a systematic coding process, into four main categories: policy, requirements, user feedback, and privacy leak analysis. Our survey revealed that the largest percentage of existing literature tackled problems related to privacy leaks, with less number of papers classified under the other three categories. Primary studies classified under each category were further summarized and their future work sections were analyzed. The analysis exposed several directions of future work in each of the categories and suggested several areas for improvement.


\ifCLASSOPTIONcaptionsoff
  \newpage
\fi

\bibliographystyle{IEEEtran}
\bibliography{tse}

\begin{thebibliography}{10}
\providecommand{\url}[1]{#1}
\csname url@samestyle\endcsname
\providecommand{\newblock}{\relax}
\providecommand{\bibinfo}[2]{#2}
\providecommand{\BIBentrySTDinterwordspacing}{\spaceskip=0pt\relax}
\providecommand{\BIBentryALTinterwordstretchfactor}{4}
\providecommand{\BIBentryALTinterwordspacing}{\spaceskip=\fontdimen2\font plus
\BIBentryALTinterwordstretchfactor\fontdimen3\font minus
  \fontdimen4\font\relax}
\providecommand{\BIBforeignlanguage}[2]{{%
\expandafter\ifx\csname l@#1\endcsname\relax
\typeout{** WARNING: IEEEtran.bst: No hyphenation pattern has been}%
\typeout{** loaded for the language `#1'. Using the pattern for}%
\typeout{** the default language instead.}%
\else
\language=\csname l@#1\endcsname
\fi
#2}}
\providecommand{\BIBdecl}{\relax}
\BIBdecl

\bibitem{Warren90}
S.~Warren and L.~Brandeis, ``The right to privacy,'' \emph{Harvard Law Review},
  vol.~4, no.~5, pp. 193--220, 1890.

\bibitem{Solove02}
D.~Solove, ``Conceptualizing privacy,'' \emph{California Law Review}, vol.~90,
  p. 1087, 2002.

\bibitem{Westin68}
A.~Westin and O.~Ruebhausen, ``Privacy and freedom,'' \emph{Washington and Lee
  Law Review}, vol.~25, p. 166, 1968.

\bibitem{Westin03}
A.~Westin, ``Social and political dimensions of privacy,'' \emph{Journal of
  Social Issues}, vol.~59, no.~2, pp. 431--453, 2003.

\bibitem{Lamsweerde01}
A.~{van Lamsweerde}, ``Goal-oriented requirements engineering: {A} guided
  tour,'' in \emph{IEEE Inter. Symposium on Requirements Engineering}, 2001,
  pp. 249--262.

\bibitem{Aydin17}
A.~Aydin, D.~Piorkowski, O.~Tripp, P.~Ferrara, and M.~Pistoia, ``Visual
  configuration of mobile privacy policies,'' in \emph{Inter. Conf. on
  Fundamental Approaches to Software Engineering}, 2017, pp. 338--355.

\bibitem{Li17}
L.~Li, T.~Bissyand{\'e}, M.~Papadakis, S.~Rasthofer, A.~Bartel, D.~Octeau,
  J.~Klein, and L.~Traon, ``Static analysis of {A}ndroid apps: {A} systematic
  literature review,'' \emph{Information and Software Technology}, vol.~88, pp.
  67--95, 2017.

\bibitem{Wang18}
X.~Wang, X.~Qin, M.~Bokaei, R.~Slavin, T.~Breaux, and J.~Niu, ``Guileak:
  {T}racing privacy policy claims on user input data for {A}ndroid
  applications,'' in \emph{Inter. Conf. on Software Engineering}, 2018, pp.
  37--47.

\bibitem{Acquisti17}
A.~Acquisti, I.~Adjerid, R.~Balebako, L.~Brandimarte, L.~F. Cranor,
  S.~Komanduri, P.~G. Leon, N.~Sadeh, F.~Schaub, M.~Sleeper, Y.~Wang, and
  S.~Wilson, ``Nudges for privacy and security: {U}nderstanding and assisting
  users' choices online,'' \emph{ACM Computing Surveys}, vol.~50, no.~3, p.~44,
  2017.

\bibitem{Papadopoulos17}
E.~Papadopoulos, M.~Diamantaris, P.~Papadopoulos, T.~Petsas, S.~Ioannidis, and
  E.~Markatos, ``The long-standing privacy debate: {M}obile websites vs mobile
  apps,'' in \emph{Inter. Conf. on World Wide Web}, 2017, pp. 153--162.

\bibitem{Kitchenham09}
B.~Kitchenham, P.~Brereton, D.~Budgen, M.~Turner, J.~Bailey, and S.~Linkman,
  ``Systematic literature reviews in software engineering--{A} tertiary
  study,'' \emph{Information and Software Technology}, vol.~51, no.~1, pp.
  7--15, 2009.

\bibitem{Wohlin12}
C.~Wohlin, P.~Runeson, M.~H\"{o}st, M.~Ohlsson, B.~Regnell, and A.~Wessl\`{e}n,
  \emph{Experimentation in Software Engineering}.\hskip 1em plus 0.5em minus
  0.4em\relax Springer, 2012.

\bibitem{Kitchenham12}
B.~Kitchenham, D.~Budgen, and P.~Brereton, ``Using mapping studies as the basis
  for further research - {A} participant-observer case study,''
  \emph{Information and Software Technology}, vol.~53, no.~6, pp. 638--651,
  2012.

\bibitem{Petersen08}
K.~Petersen, R.~Feldt, S.~Mujtaba, and M.~Mattsson, ``Systematic mapping
  studies in software engineering,'' in \emph{Inter. Conf. on Evaluation and
  Assessment in Software Engineering}, 2008, pp. 68--77.

\bibitem{Kitchenham09SLR}
B.~Kitchenham, P.~Brereton, D.~Budgen, M.~Turner, J.~Bailey, and S.~Linkman,
  ``Systematic literature reviews in software engineering – a systematic
  literature review,'' \emph{Information and Software Technology}, vol.~51,
  no.~1, pp. 7--15, 2009.

\bibitem{Wohlin14}
C.~Wohlin, ``Guidelines for snowballing in systematic literature studies and a
  replication in software engineering,'' in \emph{Inter. Conf. on evaluation
  and assessment in software engineering}, 2014, p.~38.

\bibitem{Bhatia16}
J.~Bhatia, T.~Breaux, and F.~Schaub, ``Mining privacy goals from privacy
  policies using hybridized task recomposition,'' \emph{ACM Transactions on
  Software Engineering and Methodology}, vol.~25, no.~3, p.~22, 2016.

\bibitem{Young11}
J.~Young, ``Commitment analysis to operationalize software requirements from
  privacy policies,'' \emph{Requirements Engineering}, vol.~16, no.~1, pp.
  33--46, 2011.

\bibitem{Yu18}
L.~Yu, X.~Luo, C.~Qian, S.~Wang, and H.~Leung, ``Enhancing the
  description-to-behavior fidelity in {A}ndroid apps with privacy policy,''
  \emph{IEEE Transactions on Software Engineering}, vol.~44, no.~9, pp.
  834--854, 2018.

\bibitem{Yu16}
L.~Yu, X.~Luo, C.~Qian, and S.~Wang, ``Revisiting the description-to-behavior
  fidelity in {A}ndroid applications,'' in \emph{Inter. Conf. on Software
  Analysis, Evolution, and Reengineering}, 2016, pp. 415--426.

\bibitem{Slavin16}
R.~Slavin, X.~Wang, M.~Bokaei, J.~Hester, R.~Krishnan, J.~Bhatia, T.~Breaux,
  and J.~Niu, ``Toward a framework for detecting privacy policy violations in
  {A}ndroid application code,'' in \emph{Inter. Conf. on Software Engineering},
  2016, pp. 25--36.

\bibitem{Yu18-2}
L.~Yu, X.~Luo, J.~Chen, H.~Zhou, T.~Zhang, H.~Chang, and H.~Leung,
  ``P{PC}hecker: {T}owards accessing the trustworthiness of {A}ndroid apps'
  privacy policies,'' \emph{IEEE Transactions on Software Engineering}, 2018.

\bibitem{Arzt14}
S.~Arzt, S.~Rasthofer, C.~Fritz, E.~Bodden, A.~Bartel, J.~Klein, Y.~Traon,
  D.~Octeau, and P.~McDaniel, ``Flowdroid: {P}recise context, flow, field,
  object-sensitive and lifecycle-aware taint analysis for {A}ndroid apps,''
  \emph{Acm Sigplan Notices}, vol.~49, no.~6, pp. 259--269, 2014.

\bibitem{Zimmeck16}
S.~Zimmeck, Z.~Wang, L.~Zou, R.~Iyengar, B.~Liu, F.~Shaub, S.~Wilson, N.~Sadeh,
  S.~Bellovin, and J.~Reidenberg, ``Automated analysis of privacy requirements
  for mobile apps,'' in \emph{AAAI Fall Symposium Series}, 2016, pp. 286--296.

\bibitem{Iacob13}
C.~Iacob and R.~Harrison, ``Retrieving and analyzing mobile apps feature
  requests from online reviews,'' in \emph{Mining Software Repositories}, 2013,
  pp. 41--44.

\bibitem{Jha19}
N.~Jha and A.~Mahmoud, ``Mining non-functional requirements from app store
  reviews,'' \emph{Empirical Software Engineering}, 2019.

\bibitem{Noei19}
E.~Noei, F.~Zhang, S.~Wang, and Y.~Zou, ``Towards prioritizing user-related
  issue reports of mobile applications,'' \emph{Empirical Software
  Engineering}, 2019.

\bibitem{Palomba15}
F.~Palomba, M.~{Linares-Vasquez}, G.~Bavota, R.~Oliveto, M.~Penta,
  D.~Poshyvanyk, and A.~Lucia, ``User reviews matter! {T}racking crowdsourced
  reviews to support evolution of successful apps,'' in \emph{Inter. Conf. on
  Software Maintenance and Evolution}, 2015, pp. 291--300.

\bibitem{Khalid15}
H.~Khalid, E.~Shihab, M.~Nagappan, and A.~Hassan, ``What do mobile app users
  complain about?'' \emph{IEEE Software}, vol.~32, no.~3, pp. 70--77, 2015.

\bibitem{McIlroy16}
S.~McIlroy, N.~Ali, H.~Khalid, and A.~Hassan, ``Analyzing and automatically
  labelling the types of user issues that are raised in mobile app reviews,''
  \emph{Empirical Software Engineering}, vol.~21, no.~3, pp. 1067--1106, 2016.

\bibitem{Ciurumelea17}
A.~Ciurumelea, A.~Schaufelb{\"u}hl, S.~Panichella, and H.~Gall, ``Analyzing
  reviews and code of mobile apps for better release planning,'' in
  \emph{Inter. Conf. on Software Analysis, Evolution and Reengineering}, 2017,
  pp. 91--102.

\bibitem{Scoccia18}
G.~Scoccia, S.~Ruberto, I.~Malavolta, M.~Autili, and P.~Inverardi, ``An
  investigation into {A}ndroid run-time permissions from the end users'
  perspective,'' in \emph{Inter. Conf. on Mobile Software Engineering and
  Systems}, 2018, pp. 45--55.

\bibitem{Martin17}
W.~Martin, F.~Sarro, Y.~Jia, Y.~Zhang, and M.~Harman, ``A survey of app store
  analysis for software engineering,'' \emph{IEEE Transactions on Software
  Engineering}, vol.~43, no.~9, pp. 817--847, 2017.

\bibitem{Tun12}
T.~Tun, A.~Bandara, B.~Price, Y.~Yu, C.~Haley, I.~Omoronyia, and B.~Nuseibeh,
  ``Privacy arguments: Analysing selective disclosure requirements for mobile
  applications,'' in \emph{Inter. Requirements Engineering Conf.}, 2012, pp.
  131--140.

\bibitem{Omoronyia13}
I.~Omoronyia, L.~Cavallaro, M.~Salehie, L.~Pasquale, and B.~Nuseibeh,
  ``Engineering adaptive privacy: {O}n the role of privacy awareness
  requirements,'' in \emph{Inter. Conf. on Software Engineering}, 2013, pp.
  632--641.

\bibitem{Thomas14}
K.~Thomas, A.~Bandara, B.~Price, and B.~Nuseibeh, ``Distilling privacy
  requirements for mobile applications,'' in \emph{Inter. Conf. on Software
  Engineering}, 2014, pp. 871--882.

\bibitem{Mai18}
P.~Mai, A.~Goknil, L.~Shar, F.~Pastore, L.~Briand, and S.~Shaame, ``Modeling
  security and privacy requirements: {A} use case-driven approach,''
  \emph{Information and Software Technology}, vol. 100, pp. 165--182, 2018.

\bibitem{Breaux14}
T.~Breaux, H.~Hibshi, and A.~Rao, ``Eddy, a formal language for specifying and
  analyzing data flow specifications for conflicting privacy requirements,''
  \emph{Requirements Engineering}, vol.~19, no.~3, pp. 281--307, 2014.

\bibitem{Breaux13}
T.~Breaux and A.~Rao, ``Formal analysis of privacy requirements specifications
  for multi-tier applications,'' in \emph{Inter. Requirements Engineering
  Conf.}, 2013, pp. 14--23.

\bibitem{Breaux15}
T.~Breaux, D.~Smullen, and H.~Hibshi, ``Detecting repurposing and
  over-collection in multi-party privacy requirements specifications,'' in
  \emph{Inter. Requirements Engineering Conf.}, 2015, pp. 166--175.

\bibitem{Van14}
Y.~{Van Der Sype} and W.~Maalej, ``On lawful disclosure of personal user data:
  {W}hat should app developers do?'' in \emph{Inter. Workshop on Requirements
  Engineering and Law}, 2014, pp. 25--34.

\bibitem{Liu16}
Y.~Liu and A.~Simpson, ``Privacy-preserving targeted mobile advertising:
  {R}equirements, design and a prototype implementation,'' \emph{Software:
  Practice and Experience}, vol.~46, no.~12, pp. 1657--1684, 2016.

\bibitem{kang90}
K.~Kang, S.~Cohen, J.~Hess, W.~Novak, and S.~Peterson, ``Feature-oriented
  domain analysis ({FODA}) feasibility study,'' Software Engineering Institite,
  Carnegie Mellon University, Tech. Rep. CMU/SEI-90-TR-21, 1990.

\bibitem{Bruno04}
B.~Baixauli, J.~Leite, and J.~Mylopoulos, ``Visual variability analysis for
  goal models,'' in \emph{Inter. Requirements Engineering Conf.}, 2004, pp.
  198--207.

\bibitem{Eric09}
E.~Yu, ``Social modeling and i,'' \emph{Conceptual Modeling: Foundations and
  Applications}, pp. 99--121, 2009.

\bibitem{Pohl05}
K.~Pohl, G.~B\"{o}ckle, and F.~Linden, \emph{Software Product Line Engineering:
  {F}oundations, Principles, and Techniques}.\hskip 1em plus 0.5em minus
  0.4em\relax Springer, 2005.

\bibitem{Li15Iccta}
L.~Li, A.~Bartel, T.~Bissyand{\'e}, J.~Klein, Y.~Traon, S.~Arzt, S.~Rasthofer,
  E.~Bodden, D.~Octeau, and P.~McDaniel, ``{IccTA}: {D}etecting inter-component
  privacy leaks in {A}ndroid apps,'' in \emph{Inter. Conf. on Software
  Engineering}, 2015, pp. 280--291.

\bibitem{Newsome05}
J.~Newsome and D.~Song, ``Dynamic taint analysis for automatic detection,
  analysis, and signature generation of exploits on commodity software,'' in
  \emph{Network and Distributed System Security Symposium}, 2005.

\bibitem{Demissie16}
B.~Demissie, D.~Ghio, M.~Ceccato, and A.~Avancini, ``Identifying {A}ndroid
  inter app communication vulnerabilities using static and dynamic analysis,''
  in \emph{Inter. Conf. on Mobile Software Engineering and Systems}, 2016, pp.
  255--266.

\bibitem{Enck14}
W.~Enck, P.~Gilbert, S.~Han, V.~Tendulkar, B.~Chun, L.~Cox, J.~Jung,
  P.~McDaniel, and A.~Sheth, ``Taintdroid: {A}n information-flow tracking
  system for realtime privacy monitoring on smartphones,'' \emph{ACM
  Transactions on Computer Systems}, vol.~32, no.~2, p.~5, 2014.

\bibitem{Zhang17}
Z.~Zhang and X.~Feng, ``{AndroidLeaker}: A hybrid checker for collusive leak in
  {A}ndroid applications,'' in \emph{Symposium on Dependable Software
  Engineering: Theories, Tools, and Applications}, 2017, pp. 164--180.

\bibitem{Feng14}
Y.~Feng, S.~Anand, I.~Dillig, and A.~Aiken, ``Apposcopy: {S}emantics-based
  detection of {A}ndroid malware through static analysis,'' in \emph{Inter.
  Symposium on Foundations of Software Engineering}, 2014, pp. 576--587.

\bibitem{Huang14}
J.~Huang, X.~Zhang, L.~Tan, P.~Wang, and B.~Liang, ``Asdroid: {D}etecting
  stealthy behaviors in {A}ndroid applications by user interface and program
  behavior contradiction,'' in \emph{Inter. Conf. on Software Engineering},
  2014, pp. 1036--1046.

\bibitem{Bartel14}
A.~Bartel, J.~Klein, M.~Monperrus, and Y.~Traon, ``Static analysis for
  extracting permission checks of a large scale framework: {T}he challenges and
  solutions for analyzing {A}ndroid,'' \emph{IEEE Transactions on Software
  Engineering}, vol.~40, no.~6, pp. 617--632, 2014.

\bibitem{Shen14}
F.~Shen, N.~Vishnubhotla, C.~Todarka, M.~Arora, B.~Dhandapani, E.~Lehner,
  S.~Ko, and L.~Ziarek, ``Information flows as a permission mechanism,'' in
  \emph{Inter. Conf. on Automated Software Engineering}, 2014, pp. 515--526.

\bibitem{Holavanalli13}
S.~Holavanalli, D.~Manuel, V.~Nanjundaswamy, B.~Rosenberg, F.~Shen, S.~Ko, and
  L.~Ziarek, ``Flow permissions for {A}ndroid,'' in \emph{Inter. Conf. on
  Mobile Software Engineering and Systems}, 2013, pp. 652--657.

\bibitem{Gorla14}
A.~Gorla, I.~Tavecchia, F.~Gross, and A.~Zeller, ``Checking app behavior
  against app descriptions,'' in \emph{Inter. Conf. on Software Engineering},
  2014, pp. 1025--1035.

\bibitem{Xiao15}
X.~Xiao, N.~Tillmann, M.~Fahndrich, J.~Halleux, M.~Moskal, and T.~Xie,
  ``User-aware privacy control via extended static-information-flow analysis,''
  \emph{Automated Software Engineering}, vol.~22, no.~3, pp. 333--366, 2015.

\bibitem{Huang15}
W.~Huang, Y.~Dong, A.~Milanova, and J.~Dolby, ``Scalable and precise taint
  analysis for {A}ndroid,'' in \emph{Inter. Symposium on Software Testing and
  Analysis}, 2015, pp. 106--117.

\bibitem{Yang15Appcontext}
W.~Yang, X.~Xiao, B.~Andow, S.~Li, T.~Xie, and W.~Enck, ``Appcontext:
  {D}ifferentiating malicious and benign mobile app behaviors using context,''
  in \emph{Inter. Conf. on Software Engineering}, 2015, pp. 303--313.

\bibitem{Bagheri15}
H.~Bagheri, A.~Sadeghi, J.~Garcia, and S.~Malek, ``Covert: {C}ompositional
  analysis of {A}ndroid inter-app permission leakage,'' \emph{IEEE Transactions
  on Software Engineering}, vol.~41, no.~9, pp. 866--886, 2015.

\bibitem{Avdiienko15}
V.~Avdiienko, K.~Kuznetsov, A.~Gorla, A.~Zeller, S.~Arzt, S.~Rasthofer, and
  E.~Bodden, ``Mining apps for abnormal usage of sensitive data,'' in
  \emph{Inter. Conf. on Software Engineering}, 2015, pp. 426--436.

\bibitem{Lee16}
S.~Lee, J.~Dolby, and S.~Ryu, ``Hybridroid: {S}tatic analysis framework for
  {A}ndroid hybrid applications,'' in \emph{Inter. Conf. on Automated Software
  Engineering}, 2016, pp. 250--261.

\bibitem{Li16}
L.~Li, T.~Bissyand{\'e}, D.~Octeau, and J.~Klein, ``Reflection-aware static
  analysis of {A}ndroid apps,'' in \emph{Inter. Conf. on Automated Software
  Engineering}, 2016, pp. 756--761.

\bibitem{Li16Droidra}
L.~Li, T.~Bissyand\'e, D.~Octeau, and J.~Klein, ``Droidra: {T}aming reflection
  to support whole-program analysis of {A}ndroid apps,'' in \emph{Inter.
  Symposium on Software Testing and Analysis}, 2016, pp. 318--329.

\bibitem{Meng16}
G.~Meng, Y.~Xue, Z.~Xu, Y.~Liu, J.~Zhang, and A.~Narayanan, ``Semantic
  modelling of {A}ndroid malware for effective malware comprehension,
  detection, and classification,'' in \emph{Inter. Symposium on Software
  Testing and Analysis}, 2016, pp. 306--317.

\bibitem{Rahman17}
A.~Rahman, P.~Pradhan, A.~Partho, and L.~Williams, ``Predicting {A}ndroid
  application security and privacy risk with static code metrics,'' in
  \emph{Inter. Conf. on Mobile Software Engineering and Systems}, 2017, pp.
  149--153.

\bibitem{Narayanan18}
A.~Narayanan, M.~Chandramohan, L.~Chen, and Y.~Liu, ``A multi-view
  context-aware approach to {A}ndroid malware detection and malicious code
  localization,'' \emph{Empirical Software Engineering}, vol.~23, no.~3, pp.
  1--53, 2018.

\bibitem{Zhang18}
Y.~Zhang, Y.~Li, T.~Tan, and J.~Xue, ``Ripple: {R}eflection analysis for
  {A}ndroid apps in incomplete information environments,'' \emph{Software:
  Practice and Experience}, vol.~48, no.~8, pp. 1419--1437, 2018.

\bibitem{Shan18}
Z.~Shan, I.~Neamtiu, and R.~Samuel, ``Self-hiding behavior in {A}ndroid apps:
  {D}etection and characterization,'' in \emph{Inter. Conf. on Software
  Engineering}, 2018, pp. 728--739.

\bibitem{Dilhara18}
M.~Dilhara, H.~Cai, and J.~Jenkins, ``Automated detection and repair of
  incompatible uses of runtime permissions in {A}ndroid apps,'' in \emph{Inter.
  Conf. on Mobile Software Engineering and Systems}, 2018, pp. 67--71.

\bibitem{Garcia18}
J.~Garcia, M.~Hammad, and S.~Malek, ``Lightweight, obfuscation-resilient
  detection and family identification of {A}ndroid malware,'' \emph{ACM
  Transactions on Software Engineering and Methodology}, vol.~26, no.~3, p.~11,
  2018.

\bibitem{Canfora18}
G.~Canfora, F.~Martinelli, F.~Mercaldo, V.~Nardone, A.~Santone, and
  C.~Visaggio, ``Leila: {F}ormal tool for identifying mobile malicious
  behaviour,'' \emph{IEEE Transactions on Software Engineering}, 2018.

\bibitem{Xu13}
W.~Xu, F.~Zhang, and S.~Zhu, ``Permlyzer: {A}nalyzing permission usage in
  {A}ndroid applications,'' in \emph{Inter. Symposium on Software Reliability
  Engineering}, 2013, pp. 400--410.

\bibitem{Hay15}
R.~Hay, O.~Tripp, and M.~Pistoia, ``Dynamic detection of inter-application
  communication vulnerabilities in {A}ndroid,'' in \emph{Inter. Symposium on
  Software Testing and Analysis}, 2015, pp. 118--128.

\bibitem{Keng16}
J.~Keng, ``Automated testing and notification of mobile app privacy leak-cause
  behaviours,'' in \emph{Inter. Conf. on Automated Software Engineering}, 2016,
  pp. 880--883.

\bibitem{Keng16Graph}
J.~Keng, L.~Jiang, T.~Wee, and R.~Balan, ``Graph-aided directed testing of
  {A}ndroid applications for checking runtime privacy behaviours,'' in
  \emph{Inter. Workshop on Automation of Software Test}, 2016, pp. 57--63.

\bibitem{Lee17}
Y.~Lee, J.~Bang, G.~Safi, A.~Shahbazian, Y.~Zhao, and N.~Medvidovic, ``A
  {SEALANT} for inter-app security holes in {A}ndroid,'' in \emph{Inter. Conf.
  on Software Engineering}, 2017, pp. 312--323.

\bibitem{Xu15}
H.~Xu, Y.~Zhou, C.~Gao, Y.~Kang, and M.~Lyu, ``Spyaware: {I}nvestigating the
  privacy leakage signatures in app execution traces,'' in \emph{Inter.
  Symposium on Software Reliability Engineering}, 2015, pp. 348--358.

\bibitem{Moussa17}
M.~Moussa, M.~Penta, G.~Antoniol, and G.~Beltrame, ``Accuse: {H}elping users to
  minimize {A}ndroid app privacy concerns,'' in \emph{Inter. Conf. on Mobile
  Software Engineering and Systems}, 2017, pp. 144--148.

\bibitem{Zhao18}
F.~Zhao, L.~Gao, Y.~Zhang, Z.~Wang, B.~Wang, and S.~Guo, ``You are where you
  app: {A}n assessment on location privacy of social applications,'' in
  \emph{Inter. Symposium on Software Reliability Engineering}, 2018, pp.
  236--247.

\bibitem{Sadeghi17}
A.~Sadeghi, H.~Bagheri, J.~Garcia, and S.~Malek, ``A taxonomy and qualitative
  comparison of program analysis techniques for security assessment of
  {A}ndroid software,'' \emph{IEEE Transactions on Software Engineering},
  vol.~43, no.~6, pp. 492--530, 2017.

\end{thebibliography}

\begin{IEEEbiography}[{\includegraphics[width=1in,height=1.25in,clip,keepaspectratio]{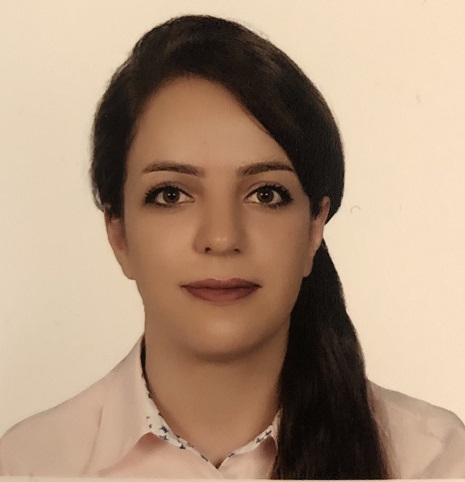}}]{Fahimeh Ebrahimi}
received the B.S. and M.S. degrees in Information Technology Engineering in 2013 and 2015 from Amirkabir University of Technology (Tehran Polytechnic). She is currently a PhD student of Computer Science and Engineering at Louisiana State University. Her main research interests include requirements engineering, app store analysis, natural language processing, and data mining. 
\end{IEEEbiography}

\begin{IEEEbiography}[{\includegraphics[width=1in,height=1.25in,clip,keepaspectratio]{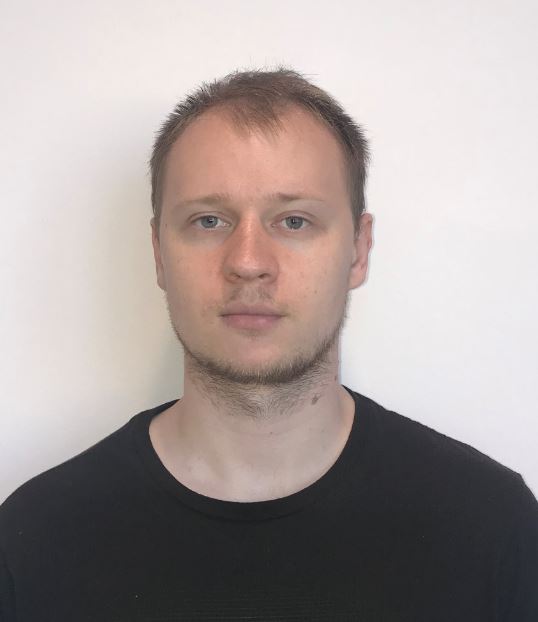}}]{Miroslav Tushev}
received the B.S. in management and organization in 2013 from the Russian Academy of National Economy and Public Administration. He is currently a PhD student of Computer Science and Engineering at Louisiana State University. His main research interests include open source systems, program comprehension, code navigation, and natural language analysis of software. 
\end{IEEEbiography}

\begin{IEEEbiography}[{\includegraphics[width=1in,height=1.25in,clip,keepaspectratio]{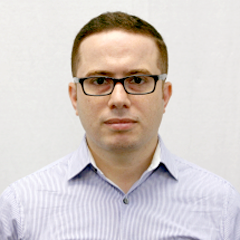}}]{Anas Mahmoud}
received the M.S. and PhD degrees in Computer Science and Engineering in 2009 and 2014 from Mississippi State University. He is
currently an assistant professor of Computer Science and Engineering at Louisiana State University. His main research interests include requirements engineering, software testing, program comprehension, code navigation, natural language analysis of software, program refactoring and information foraging. 
\end{IEEEbiography}

\end{document}